\documentclass[intlimits,twoside,a4paper]{article}

\usepackage{amsmath,amssymb}
\usepackage{graphicx}

\usepackage[T2A]{fontenc}
\usepackage[cp1251]{inputenc}
\usepackage{color}

\usepackage[eqsecnum]{cmpj3}

\issue{2018}{21}{4}{43301}
\doinumber{10.5488/CMP.21.43301}

\title[Investigation on stilbene molecular system device]%
{Switching behaviour of stilbene molecular device: a first-principles study}%
\author[V. Nagarajan, R. Chandiramouli]{V. Nagarajan, R. Chandiramouli\footnote{Corresponding author} }
\address{
School of Electrical and Electronics Engineering,
SASTRA Deemed University, \\Tirumalaisamudram, Thanjavur --- 613 401, India
}

\date{Received May 21, 2018, in final form August 29, 2018}

\DeclareMathOperator{\Tr}{Tr}

\begin{document}

\maketitle

\begin{abstract}
The switching behaviour of stilbene molecular system (SMS) device is investigated with the help of non-equilib\-rium Green’s function (NEGF) approach using first principles calculation. The transmission spectrum of cis-isomers confirmed that more electrons are transferred across the SMS-device using optical excitation by the spin of C$=$C bond by torsion angle $(\theta=180^\circ)$. The current-voltage characteristics show the lower magnitude of current for trans-stilbene and higher magnitude of current for cis-stilbene for an externally applied bias voltage. The outcome of the proposed work suggests that cis and trans-stilbene molecular device can be used as a switch.
\pacs 33.15.Mt, 61.43.Bn, 61.46.Bc, 78.20.Bh 
\end{abstract}

\section{Introduction}

Photochromes, generally represented as molecular switches, allow tuning their functional properties, which arose due to their molecular structure through excitation of a light source of suitable wavelengths. Thus, the molecules can be inter-converted reversibly among the two metastable isomers as a consequence of photocyclization, i.e., cis$\rightleftharpoons$trans isomerization or combination of both. Among the most significant photochromes, azobenzene, stilbene, and their derivatives exhibit reversible photo-induced cis$\rightleftharpoons$trans isomerization. Moreover, thermodynamically trans-isomer is more stable than cis form \cite{1}. Optical excitation developed by rotation of N$=$N (C$=$C) double bond in azobenzene (stilbene) or C inversion in stilbene drives the photoisomerization. Azobenzene isomers and its derivatives can be thermally initiated for cis$\rightarrow$trans  reaction. However, in the case of stilbenoid isomers, the thermal-back reaction is not feasible, because  the activation barrier is noticed to be higher by the order of 1.9~eV \cite{2} compared to azobenzenes (1~eV) \cite{3}. The switching behaviour of stilbene highly depends on the structure of both stilbene conformations (cis and trans). The switching properties of stilbene from trans to cis switching is observed by ultraviolet irradiation  (340--400~nm), whereas cis form to trans switching is noticed with the exposure of visible radiation (400--500~nm)  \cite{2}. The lifetime for excitation of cis and trans form is noticed to be around 300~fs and 10--100~ps, respectively. In addition, the energy barrier of cis and trans conformations in the gas phase has been found experimentally to be $<0.05$~eV and 0.15~eV, respectively.

Nowadays, much effort has been taken by researchers toward molecular switches owing to their potential application in nanotechnology including information storage and developing adaptive surfaces \cite{4}. The primary goal of the future applications is to control and understand the geometrical variations of molecular switches, such as stilbenes and azobenzenes adsorbed on the surface of a solid substrate. For this reason, it is essential to gather a detailed knowledge regarding the electronic structure (occupied and unoccupied orbitals) and the adsorption structure (molecular orientation) of the molecules in contact with solid substrates. Many researchers, particularly, Kaloni group, have  extensively studied the structural and electron properties of organic polymers, especially polythiophene, to be utilized for future optical and electronic devices \cite{5,6,7,8,9,10}.

Photochromic molecules can experience reversible photo-triggered isomerization among the two meta-stable states. Recently, a larger quantity of photochromic molecules have been synthesized and designed, including E/Z isomerization, cycloadditions, valence isomerization and tautomerizations. Usually, the photochromic molecules can be segregated into numerous classes based on their type of chemical process concerned \cite{11}: (1) photo-induced bond cleavages, namely perchlorotoluene; (2) pericyclic reactions, such as electrocyclizations, such as oxazines/-spiropyrans, cycloaditions, diarylethenes and fulgides in aromatic compounds; (3) electron transfers; (4) intramolecular group/hydrogen transfer such as polycyclic and anilsquinones; (5) E/Z isomerizations including stilibenes, azobenzenes and so on. Among these photochromic molecules, the extensively used are stilbenes \cite{12}, diarylethenes \cite{13}, azobenzenes \cite{1} and oxazines/-spiropyrans \cite{14}. Moreover, oxazines/-spiropyrans, stilbenes and azobenzenes can be switched between the possible two isomers either thermally or photochemically, whereas the diarylethenes can be reverted either electrochemically or photochemically. The functionalization of carbon-dependent materials with photochromic molecules is an extensively used technique to observe their capability to respond towards light stimuli at particular wavelengths. Furthermore, the photochromic isomers adopt various significant properties on a single molecule, which leads to considerably diverse macroscopic properties at the ensemble level \cite{15}. Thus, the decoration of carbon-dependent materials along with photochromic moieties to regulate local changes in the electrostatic, mechanical, optical environment through a light source input has shown great interest in the subject of materials science, physics and chemistry.

Leyssner et al. \cite{16} reported the molecular switches on Au(111) electrodes and compared the switching properties of stilbene and azobenzene derivatives. Zhang and co-workers \cite{17} proposed the optical response of photochromic molecules with the coupling of carbon nanomaterials. Gutierrez et al. \cite{18} reported the transport properties of cis/trans stilbenoid with a carbon-based molecular system using DFT technique. DFT is a significant technique to explore the electronic properties of an organic molecule, which can be investigated at the atomistic level. The inspiration behind the proposed study is to investigate the molecular switching behaviour of trans and cis isomers of stilbene, which are sandwiched amongst Au(111) electrodes in order to investigate the transport characteristics.

\section {Computational methods}
In this work, the electron transport property of SMS device is investigated using NEGF and DFT approach executed in TranSIESTA utility in SIESTA package \cite{19}. So far, the NEGF technique has been utilized to explore the coherent transport via molecular device \cite{20}. It infers that the quantum-mechanical-phase provides a coherent transfer of electrons between two electrodes, via the molecular region. Nevertheless, the NEGF approach is also a promising technique for phonon scattering and determining the external parameters, magnetic field for instance. All structural optimization in the present work is carried out by adjusting the grid mesh cut-off and vacuum slab about 550~eV and 16~\AA, respectively. The full geometrical optimization was performed with the help of conjugate gradient technique until the Hellmann-Feynman force was perfectly converged to $0.02~\text{eV}/\text{\AA}$. In addition, the integration of Brillouin-zone for a stilbene molecular system is sampled with ($10 \times 10 \times 10$) $\Gamma$-centered Monkhorst-pack $k$-grid \cite{21}. The electron-electron interaction on SMS-device is determined using a prominent GGA/PBE exchange correlation functional \cite{22, 23}. TranSIESTA also explores the switching mechanism of the stilbene molecular device with Troullier-Martins (TM) norm-conserving pseudopotentials scheme \cite{24}. In the present study, double zeta polarization (DZP) basis set is used for Au(111) electrode and for stilbene molecules, during structural relaxation \cite{25, 26}. The electronic transport characteristics of a stilbene molecular system are studied using a two-probe method \cite{27}. In the present model, the electrode consists of three Au(111) layers. In this stilbene molecule device, we adopted a supercell size of $3 \times 3$ with a lattice constant of 19.634~\AA. The stilbene 2-probe system is a bare system, the right-hand and left-hand gold-electrode region comprises three layers of gold (111) with $3 \times 3$ surface atoms, that replicate periodically, which forms the infinite Au-electrode. The stilbene-electrode distance is fixed to be  constant for all the system after the energy convergence of the stilbene molecular system, the axial distance of Au-S  varies from 1 to 5 with a minimum step variation of 0.1~\AA. Finally, the total energies of the stilbene molecule scattering region (MSR) for a 3-supercell system are computed, and the minimum value is observed at 2.35~{\AA} for both Au-left-hand and Au-right-hand electrodes. The methods are similar to those used in the literature \cite{28}, which use nanostructures such as carbon nanotubes and boron nanotubes. Along the stilbene molecular device, an external bias voltage ($V_\text{b}$) is applied along the right-hand-Au and left-hand-Au electrodes for the current flow.

\section {Results and discussion}
\subsection {Structure of SMS device}

\begin{figure}[!b]
\centering
\includegraphics[width=0.65\textwidth]{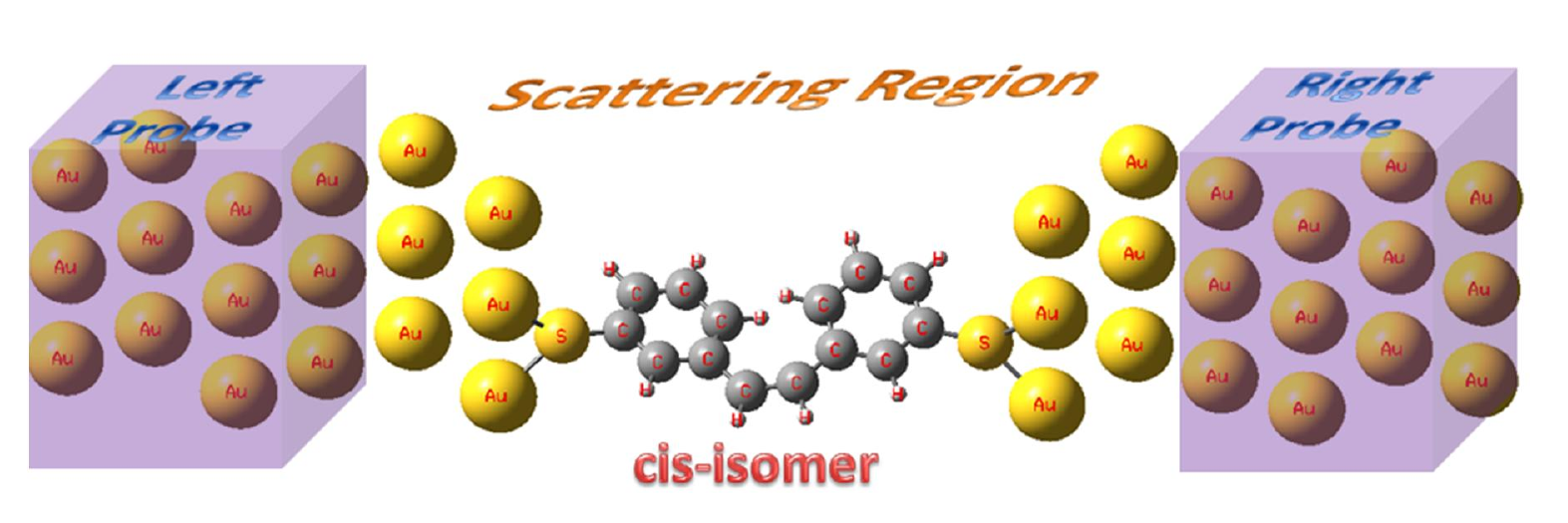}
\caption{(Colour online) The pictorial representation of cis-stilbene molecular device $(\theta=180^\circ)$.} \label{fig-s1}
\end{figure}
\begin{figure}[!b]
\centerline{\includegraphics[width=0.65\textwidth]{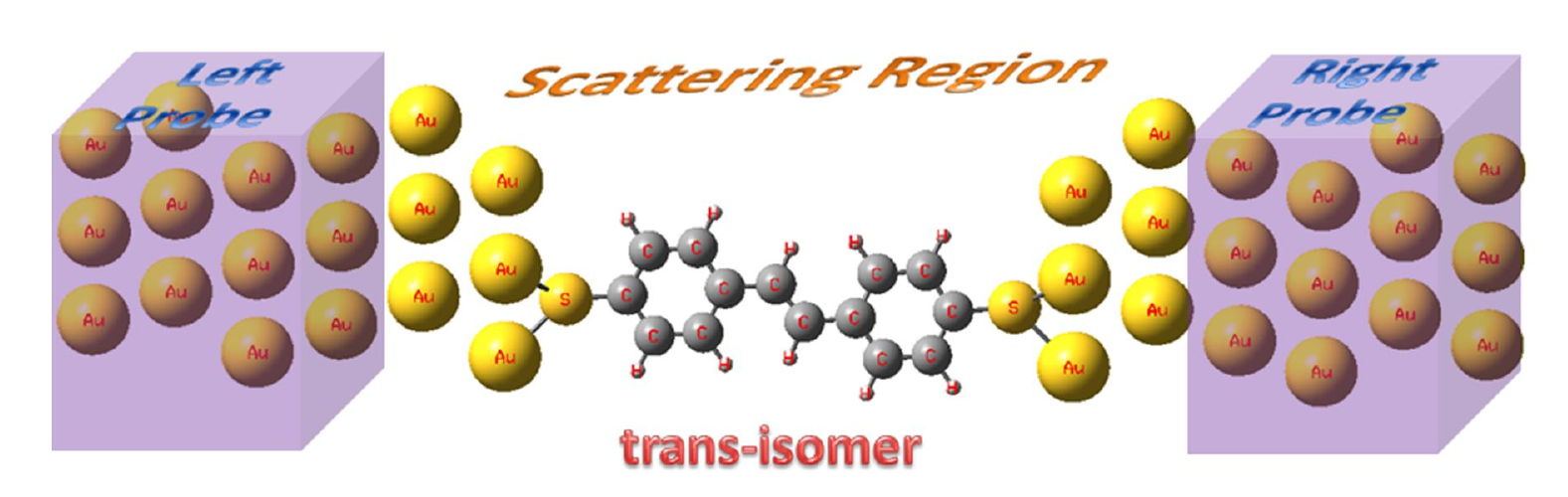}}
\caption{(Colour online) The pictorial representation of trans-stilbene molecular device $(\theta=0^\circ)$.} \label{fig-s2}
\end{figure}

The two conformers of stilbene, namely ‘trans’ and ‘cis’, are bonded to the gold electrodes. The stilbene photochromic molecules are bonded with gold electrodes through sulphur atoms. In general, noble metals, namely platinum or gold, can be utilized as electrodes, which also possess a good ohmic contact. Further, the Au-electrode is oriented along (111) plane. 
To investigate the transport characteristics of stilbene based two-probe device, DFT with NEGF method is used as demonstrated by Brandbyge and co-workers \cite{29}. The method mainly depends on the two-probe device under applied bias condition. In the present study, the two-probe stilbene device consists of two Au-electrodes on both left and right sides of the scattering region and central scattering region, which include a small portion of the Au-electrodes and stilbene molecule. At the initial stage, the electronic properties of both right-hand and left-hand Au-electrodes are computed to get a corresponding self-consistent potential (SCP), which offers the real-space boundary (RSB) conditions to the stilbene molecule scattering region (MSR). Moreover, from the Green’s function (GF) of stilbene MSR, the density matrix along with electron density is found. This stepwise procedure is repeated until the desired SCP is achieved. Further, the current flowing along the stilbene MSR can be determined from the corresponding self-energies and GF by Landauer-B\"uttiker formalism for the proposed stilbene molecular device. The stilbene molecule has properly merged to two similar semi-infinite Au(111) electrodes, as shown in figure~\ref{fig-s1}. For both isomers, namely trans and cis of stilbene, the current transferring over the SR is investigated within the voltage range of 0.1--1.2~V. Besides, for higher bias voltages, the stilbene SR gets damaged resulting in an open circuit. Therefore, the applied external $V_\text{b}$ is limited to 1.2~V and the response is conversed in the present work. The potential difference between Au-left-hand ($+V/2$) and Au-right-hand ($-V/2$) is kept across the stilbene molecular system. Interestingly, the changes in geometry and the applied external bias voltage of photochromic organic molecules influence the charge transmission and density of states (DOS) spectrum along SMS device. Figures~\ref{fig-s1} and \ref{fig-s2} illustrate the pictorial representation of SMS switching device in ON $(\theta=180^\circ)$ state and OFF $(\theta=0^\circ)$ state, respectively.

\subsection {Device density of states (DDOS) spectrum of SMS device}

\begin{figure}[!b]
\centerline{\includegraphics[width=0.95\textwidth]{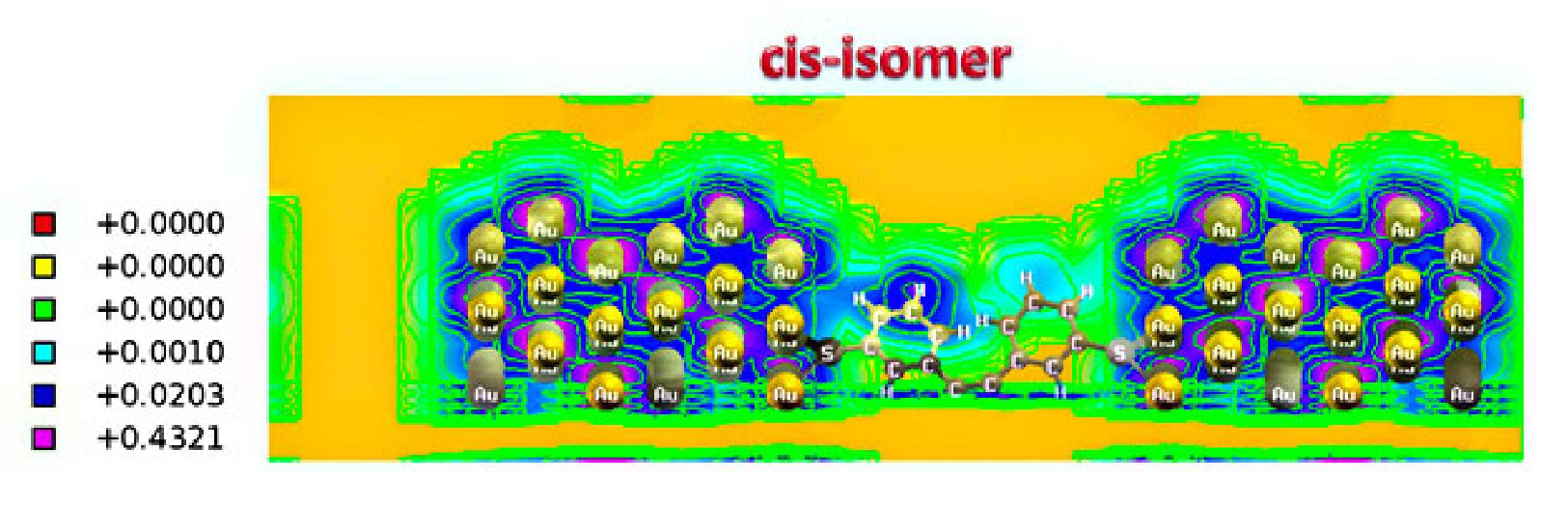}}
\caption{(Colour online) The electron density of cis-stilbene molecular device.} \label{fig-s3}
\end{figure}

\begin{figure}[!t]
\centerline{\includegraphics[width=0.95\textwidth]{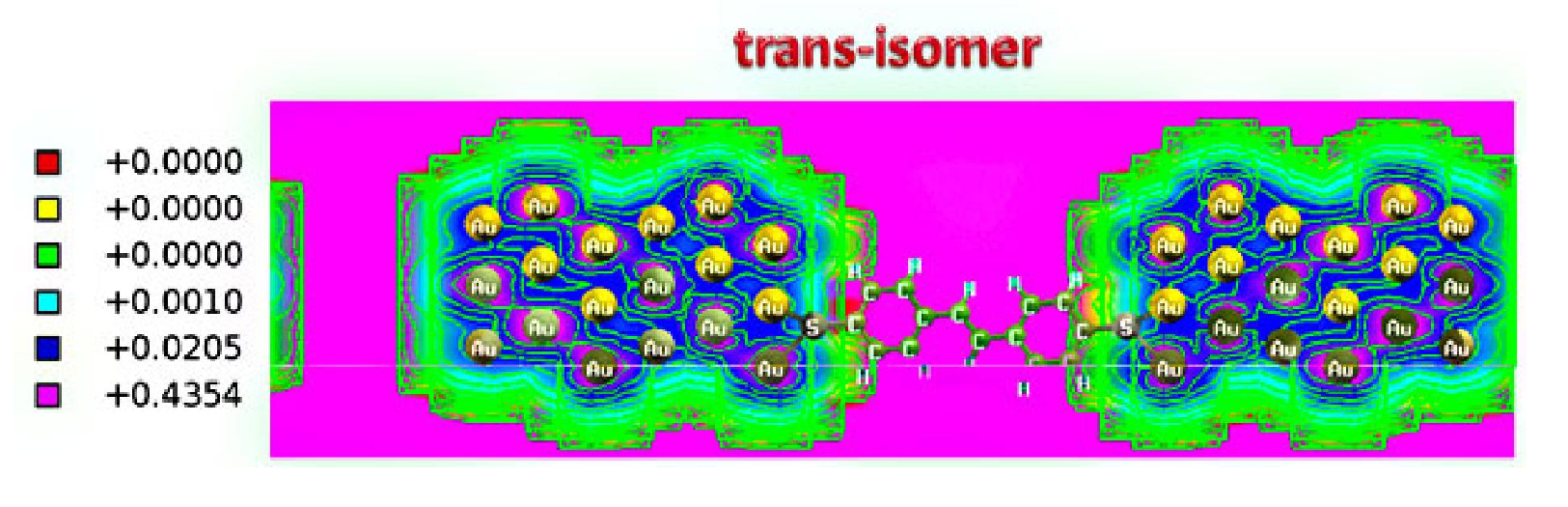}}
\caption{(Colour online) The electron density of trans-stilbene molecular device.} \label{fig-s4}
\end{figure}

To understand the electronic properties of stilbene in depth, DOS spectrum gives a clear picture about the charge density for various energy intervals under the applied external bias voltage \cite{30, 31}. Moreover, stilbenoid and stilbenes organic compounds exhibit photochemical cis-trans isomerization. Thus, stilbene also serves as building blocks for organic compounds, whose properties can be utilized in electro-optics and optical applications including nonlinear optics, optical data storage \cite{32}. The DDOS spectrum provides a perception on charge localization in SMS-device. Furthermore, based on the conformers (cis or trans), the charge localization varies upon the applied external bias voltage. The switching properties of stilbene are mainly related to the geometry of each stilbene organic molecules. The switching behaviour of stilbene from cis to trans form is observed by the exposure of visible radiation (400--500~nm) \cite{33}, whereas trans form to cis switching is noticed with the ultraviolet irradiation (340--400~nm) \cite{2}. The main objective of the present study is to model the photoisomerization along with its response coordinate described by the torsion angle around the central C$=$C bond \cite{34}. In general, isomerization mechanism may be observed from two significant models: (1) conventional one-bond flip (OBF) model, the substituents of the C$=$C bond transfer along the cone surface, (2) hula-twist (HT) \cite{34} model, only the C$=$H group rotates out of plane and the remaining parts reorient within the plane. After excitation, the trans isomer transforms to cis stilbene upon the shining ultraviolet radiation, the C$=$C bond twists by $180^\circ$, and an excitation energy ($S_1$) minimum at a normal conformer $(P^\ast)$ is observed for both cis and trans isomers \cite{35}. In order to leave the fluorescent region on both excitation $\text{trans}^\ast$ and $\text{cis}^\ast$ isomers towards $P^\ast$, the organic molecules should overcome a very small barrier. The lifetime for excitation of trans population is observed between 10 to 100~ps based on the excess energy \cite{36} and for cis photo-exited population 300~fs \cite{37}. Further, the energy barrier of an initial geometry of trans and cis stilbene in the gas phase has been found experimentally about 0.15~eV \cite{38} and $<0.05$~eV \cite{39}, respectively. The switching behaviour along the stilbene arises owing to the distribution of free electrons in C$=$C, the free electrons in two sides of trans isomer transfer to one side in cis isomer upon the exposure to the radiation of ultraviolet light. Precisely, the two $\piup$-conjugated arms of stilbene by-products participate  in the switching amongst cis and trans isomers. Interestingly, the variation in electronic structures is noticed to be significant on both trans and cis photoisomerization. Therefore, the free lone pair of electrons of C$=$C is engaged in the variation of DOS. The electron density (ED) of cis and trans SMS-device at zero bias condition is displayed in figures~\ref{fig-s3} and \ref{fig-s4}, respectively.

\begin{figure}[!b]
\vspace{-3mm}
\centerline{\includegraphics[width=0.625\textwidth]{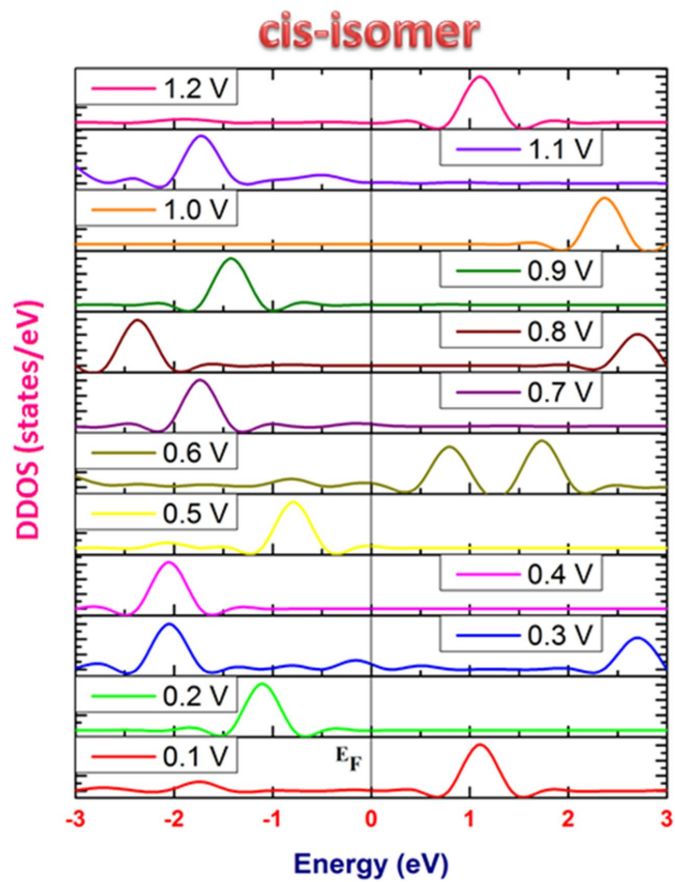}}
\caption{(Colour online) The device density of states (DDOS) spectrum of cis-stilbene molecular device.} \label{fig-s5}
\end{figure}
\begin{figure}[!t]
\centerline{\includegraphics[width=0.625\textwidth]{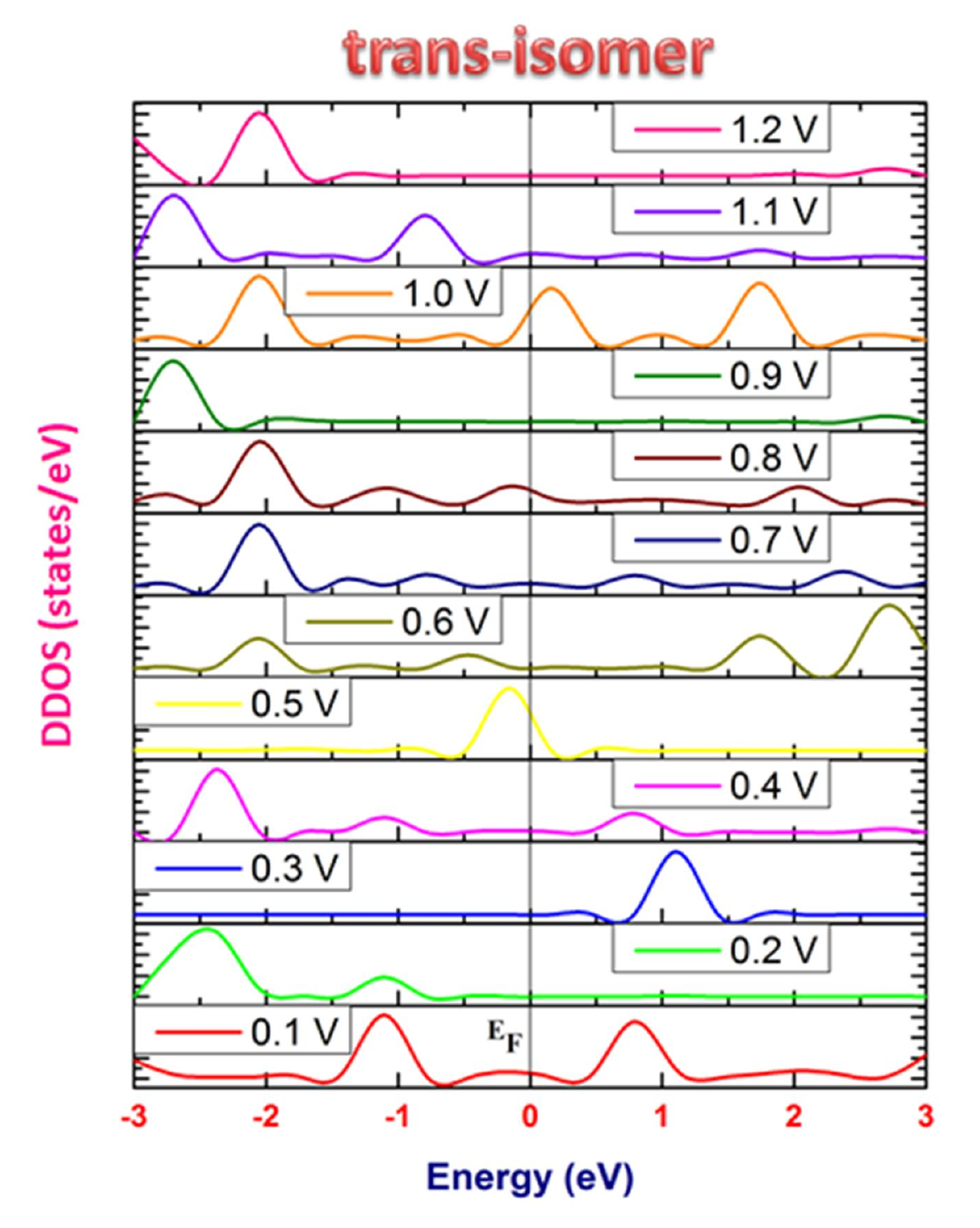}}
\caption{(Colour online) The device density of states (DDOS) spectrum of trans-stilbene molecular device.} \label{fig-s6}
\end{figure}

The electron density is noticed to be high beside the Au-electrodes regions owing to its metallic-nature. Nevertheless, in SR-region, the electron density is detected to be low. Furthermore, on changing the external $V_\text{b}$ through the $\piup$-conjugated isomers, the ED varies upon the applied external $V_\text{b}$. Besides, the changes in $V_\text{b}$ along SMS device result in the variation of the charge density across energy intervals. In this work, the changes in DOS spectrum are noticed only beyond an external bias voltage value of 0.1~V, which is proved through the considerable changes in the charge density. Moreover, a significant change in output current is observed only within the $V_\text{b}$ range of 0.1 to 1.2~V; beyond 1.2~V, the stilbene MSR gets damaged. Moreover, the Fermi level energy $(E_\text{F})$ is kept at zero, the bias window among Au-left-hand and Au-right-hand electrode is adjusted to $+V/2$ and $-V/2$ in a stilbene molecular system. Figures~\ref{fig-s5} and \ref{fig-s6} refer the DOS-spectrum of SMS device in ON (cis) and OFF (trans) states, respectively. Diligently, the orbital overlapping of stilbene hydrocarbons bonded between phenyl rings and Au atoms results in a peak maximum (Pmax) and is detected on both the valence band (VB) and on the conduction band (CB). Moreover, the switching between cis and trans isomers is noticed upon the exposure to UV radiation due to a bond flip process across C$=$C, with twisting of phenyl rings to $180^\circ$ along the surface of a cone. Besides, when the SMS device is in ON (cis) state, the Pmax is observed near CB. When $V_\text{b}$ is set to 0.1~V, the Pmax is noticed around 1.12~eV at CB. Upon increasing the external bias voltage between the Au-left-hand and Au-right-hand electrodes, the Pmax are detected in both CB as well as in VB with a peak shift owing to the variation in the highest occupied molecular orbital (HOMO) and the lowest unoccupied molecular orbital (LUMO). Furthermore, for higher voltages, peak maxima are noticed at higher energy intervals on both VB and CB. It is inferred that in cis isomers of a stilbene molecular device, the external bias voltage leads to a charge  delocalization in various energy intervals.

The variation in HOMO-LUMO orbitals along the SMS device arises owing to the one-bond flip mechanism of C$=$C bond in phenyl ring with respect to the Au-electrodes. Further, the variation in LUMO and HOMO molecular orbitals is observed at higher bias voltages, which infers that the applied external $V_\text{b}$ drives the density of charge along the SMS-device. For trans-isomers (OFF) condition, the charge density is less, which is revealed from the decrease in peak maxima in CB of DOS-spectrum as shown in figure~\ref{fig-s6}.

The torsion barrier for trans-isomers is reported to be around 3~kcal/mol \cite{40}.  In the present work, more peaks are detected and the peak shift in different energy intervals is noticed in the conduction band of cis-isomers (ON) state of SMS device rather than of trans-isomers (OFF) state. Therefore, the charge density along the stilbene molecular system changes with the help of the applied $V_\text{b}$ and the spin of the photochromic molecule through a desired angle in different conformations.

\subsection {Transport properties of SMS device}

The electronic transport characteristics of SMS device can be illustrated with the help of transmission spectrum (TS) \cite{41,42,43} The transport properties of SMS device are explored using TranSIESTA utility in SIESTA code. The current flowing through the stilbene molecular device is given by Landauer-B\"uttiker formula \cite{44}
\begin{equation}
\label{1}
I= \frac{2e}{h} \int_{+\infty}^{-\infty} T \left(E,  V_\text{b} \right) \left[f_\text{L} \left(E - \mu_\text{L}\right) - f_\text{R} \left(E - \mu_\text{R}\right)   \right]  \rd E,	
\end{equation}
\begin{equation}
\label{2}
T (E, V_\text{b}) = \Tr [ \Gamma_\text{L} (E) G^\text{R} (E) \Gamma_\text{R} (E) G^\text{A} (E)],
\end{equation}
where $\mu_\text{L}$ and  $\mu_\text{R}$ illustrates the chemical potential (CP) of Au-left-hand and Au-right-hand electrodes, respectively. $f_\text{L}$ and $f_\text{R}$ refer to the electronic Fermi functions of the Au-left-hand and Au-right-hand electrodes, respectively. $V_\text{b}$ is the applied external bias voltage of Au-left-hand and Au-right-hand electrodes \cite{45}. $T (E,V_\text{b} )$ refers to transmittance of a stilbene molecular system at the bias voltage $V_\text{b}$ and energy level $E$ \cite{46}. $\Gamma_\text{L,R}$ is referred to as a coupling function of left-hand and right-hand self-energies, respectively. The corresponding retarded and advanced Green’s function has been utilized to find the transmission of the stilbene molecular device and it is denoted by $G^\text{R}$ and $G^\text{A}$. The expansion of each term including transmittance and CP on the above equations~(\ref{1}) and (\ref{2}) is clearly discussed in our previously reported work \cite{1, 30}

 \begin{figure}[!b]
\centerline{\includegraphics[width=0.65\textwidth]{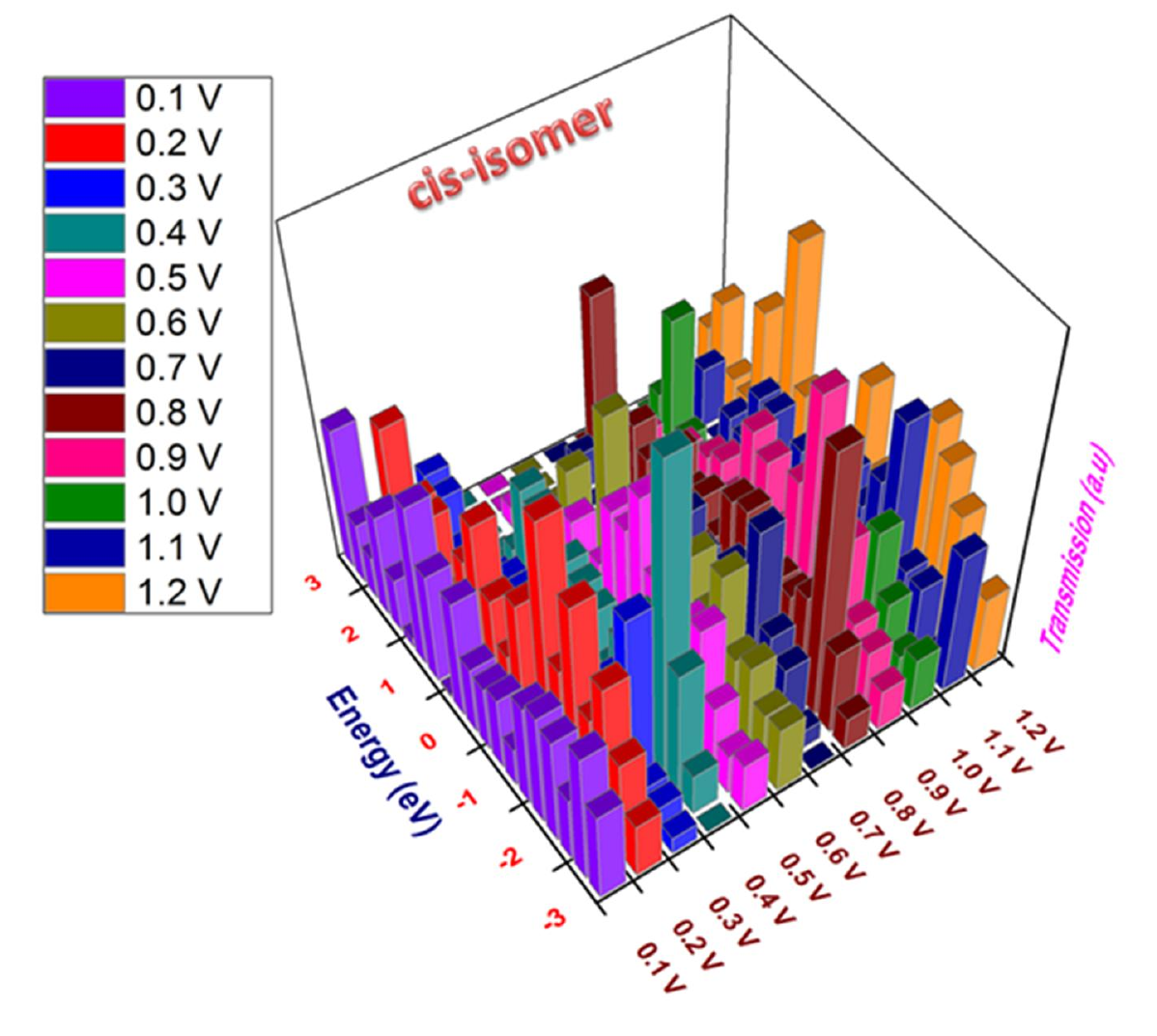}}
\caption{(Colour online) The transmission spectrum of stilbene device in cis state.} \label{fig-s7}
\end{figure}

\begin{figure}[!t]
\centerline{\includegraphics[width=0.65\textwidth]{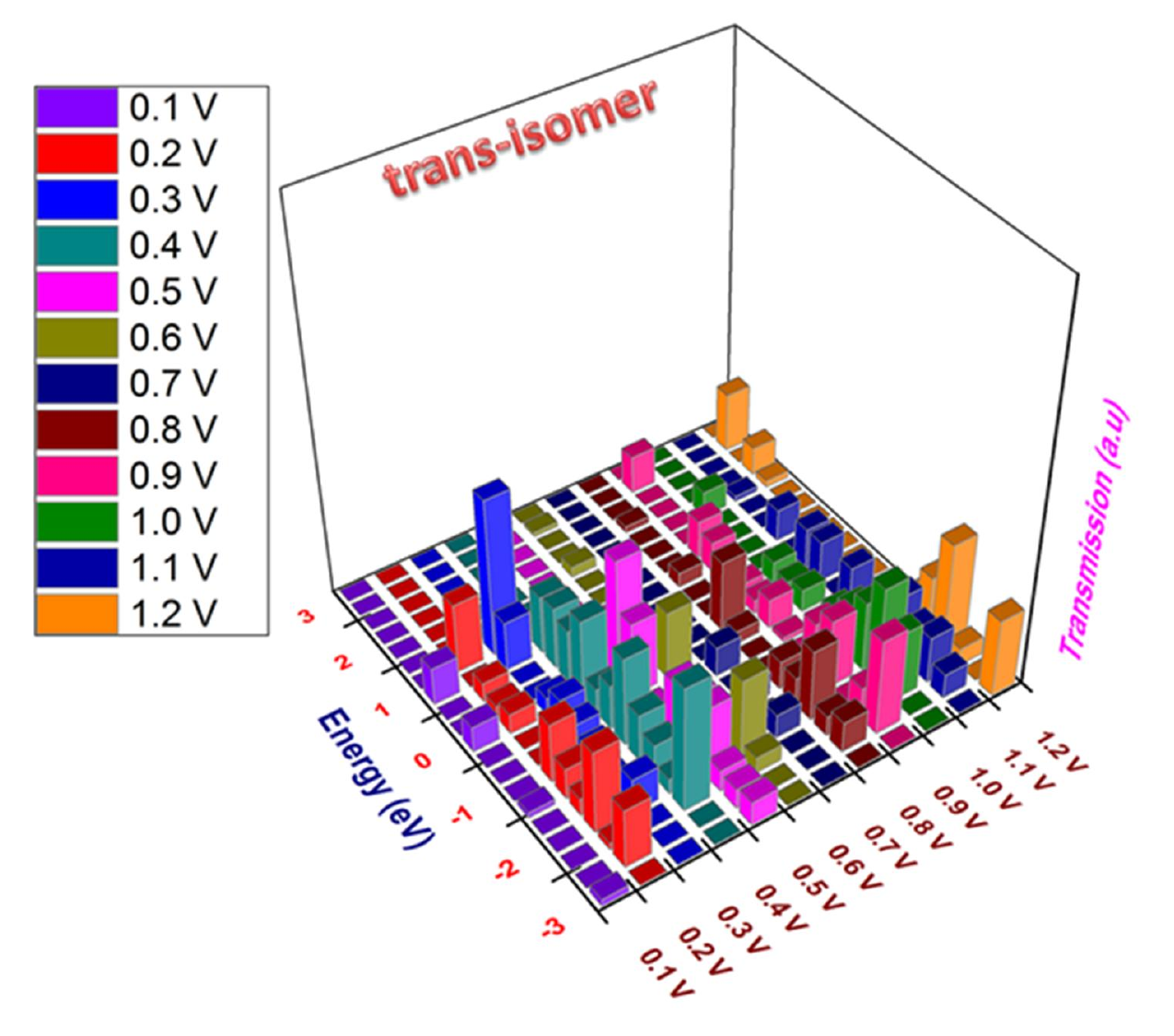}}
\caption{(Colour online) The transmission spectrum of stilbene device in trans state.} \label{fig-s8}
\end{figure}

If $V_\text{b}$ is set to zero voltage, the transmission energy at Fermi level $T(E_\text{F})$ provides the conductance $G=G_0$ $T(E_\text{F})$, where $G_0$ = $(2e^2)/h$ is the unit of quantum conductance \cite{29, 47}. 
The orbital delocalization along the stilbene molecular system directs to enhance the concentration of free electrons, where it is noticed on particular peak amplitudes $P_\text{amp}$ in the transmission spectrum of SMS device \cite{48,49,50,51}. The conductance $G$ is explored for various lengths of Au-electrodes. The conductance of stilbene at $E_\text{F}$ decreases exponentially as the gold electrodes length increases. An exponential decrease of conductance with an increasing length of an electrode has been reported in the literature \cite{52}.  Therefore, a suitable length of Au-electrode must be selected to explore the transport characteristics of a stilbene device. Further, the applied external $V_\text{b}$ acts as a driving force along the stilbene device, which results in the shift of the peak amplitudes on both CB and VB. Figures~\ref{fig-s7} and \ref{fig-s8} refer to the transmission spectrum of SMS device in cis state and trans state, respectively. The variations in the geometrical structure introduce the deviation in molecular dipole. Moreover, for cis-SMS device, for the applied external $V_\text{b}$, the $P_\text{amp}$ is noticed in VB and CB with a higher amplitude. In addition, an increase in $V_\text{b}$ pushes the charges through the stilbene molecule, that is why the peak shifts are noticed in the transmission spectrum of cis-SMS device. Nevertheless, observing the transmission spectrum of trans-stilbene device, almost all the peaks are noticed nearer to the Fermi energy level and in the valence band with lower peak amplitudes.

From the observation, it is precisely inferred that a greater peak shift and high amplitudes are noticed in CB of $180^\circ$ phase variation of SMS device rather than in $0^\circ$ phase shift of SMS device. The photoisomerization can be stimulated by optical excitation of a prominent wavelength of visible radiation (cis $\rightarrow$ trans) and UV radiation (trans $\rightarrow$ cis) owing to C inversion. Thus, it is evident that the concentration of electrons across the cis-stilbene is high, which facilitates the electronic transport across the scattering region.  Furthermore, the charge transmission in different energy intervals upon increasing $V_\text{b}$ is noticed via the shift in $P_\text{amp}$.  From the outcome, it is clearly proved that the variation in transmission spectrum along the stilbene molecular system device arose due to the one-bond flip mechanism of C$=$C bond in the phenyl ring. For trans-isomers, the transfer of electrons across the stilbene junction is less, which is proved by a decrease of peak amplitude on both VB and CB of the transmission spectrum as depicted in figure~\ref{fig-s8}. Moreover, the variation in the transmission spectrum between the cis and trans conformers is governed by the fact that there is a  potential barrier of around 3 kcal/mol \cite{40} for torsion angle $(\theta=0^\circ)$ in trans-isomer. On the other hand, there is a much less barrier noticed for cis-isomer \cite{39}. From the observation of transmission spectrum, it is precisely proved that the $V_\text{b}$ applied across the phenyl ring with a desired torsion angle governs the transition of electrons along SMS device.

\subsection {Current-voltage characteristics of stilbene molecular device}

\begin{figure}[!t]
\centerline{\includegraphics[width=0.63\textwidth]{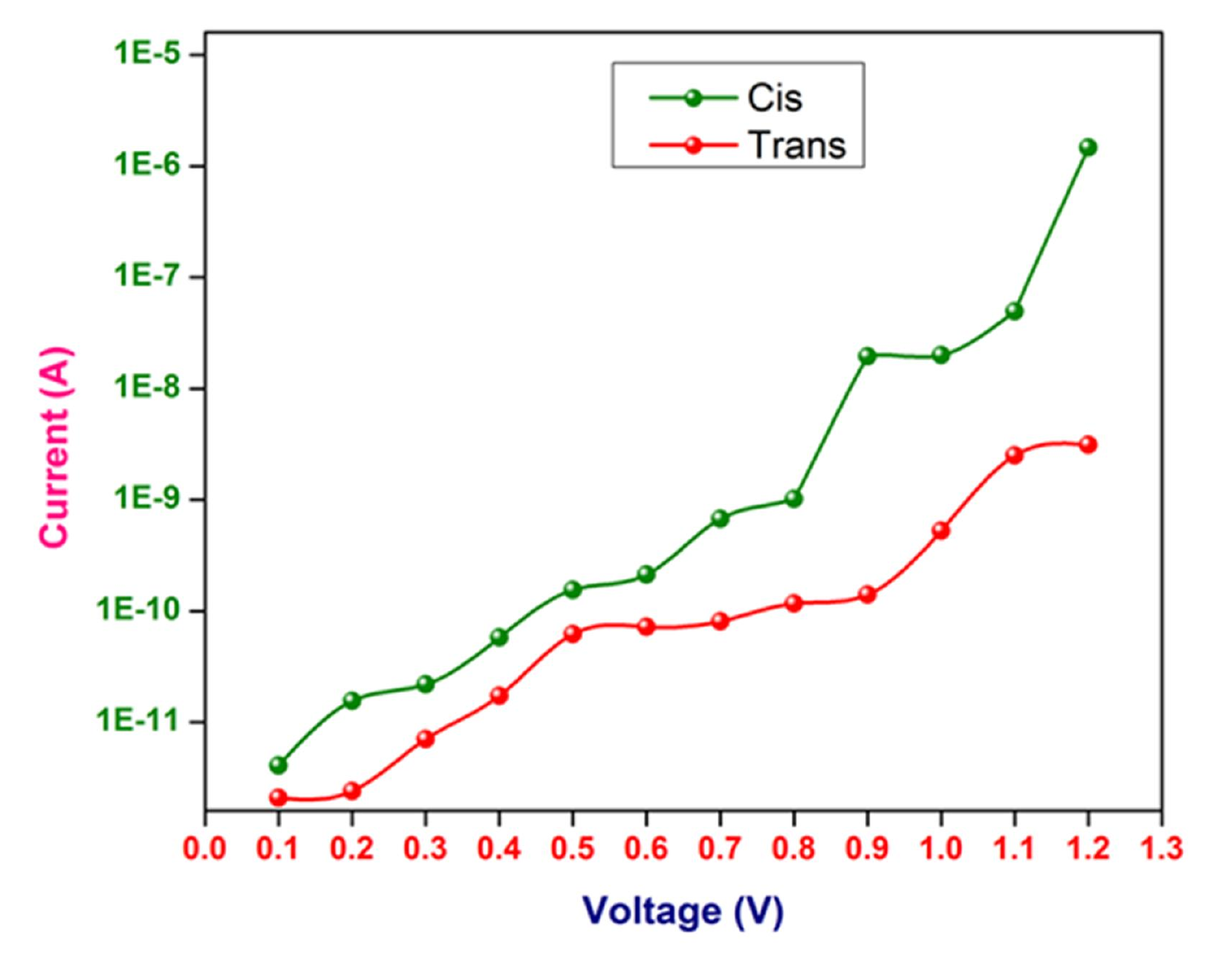}}
\vspace{-3mm}
\caption{(Colour online) The $I{-}V$ characteristics of trans and cis isomers of stilbene device.} \label{fig-s9}
\end{figure}

The $I{-}V$ characteristics provide a perception on the switching properties of cis-SMS and trans-SMS device. Figure~\ref{fig-s9} refers to the $I{-}V$ characteristics of trans and cis isomers of stilbene device.

\begin{figure}[!t]
\centerline{\includegraphics[width=0.65\textwidth]{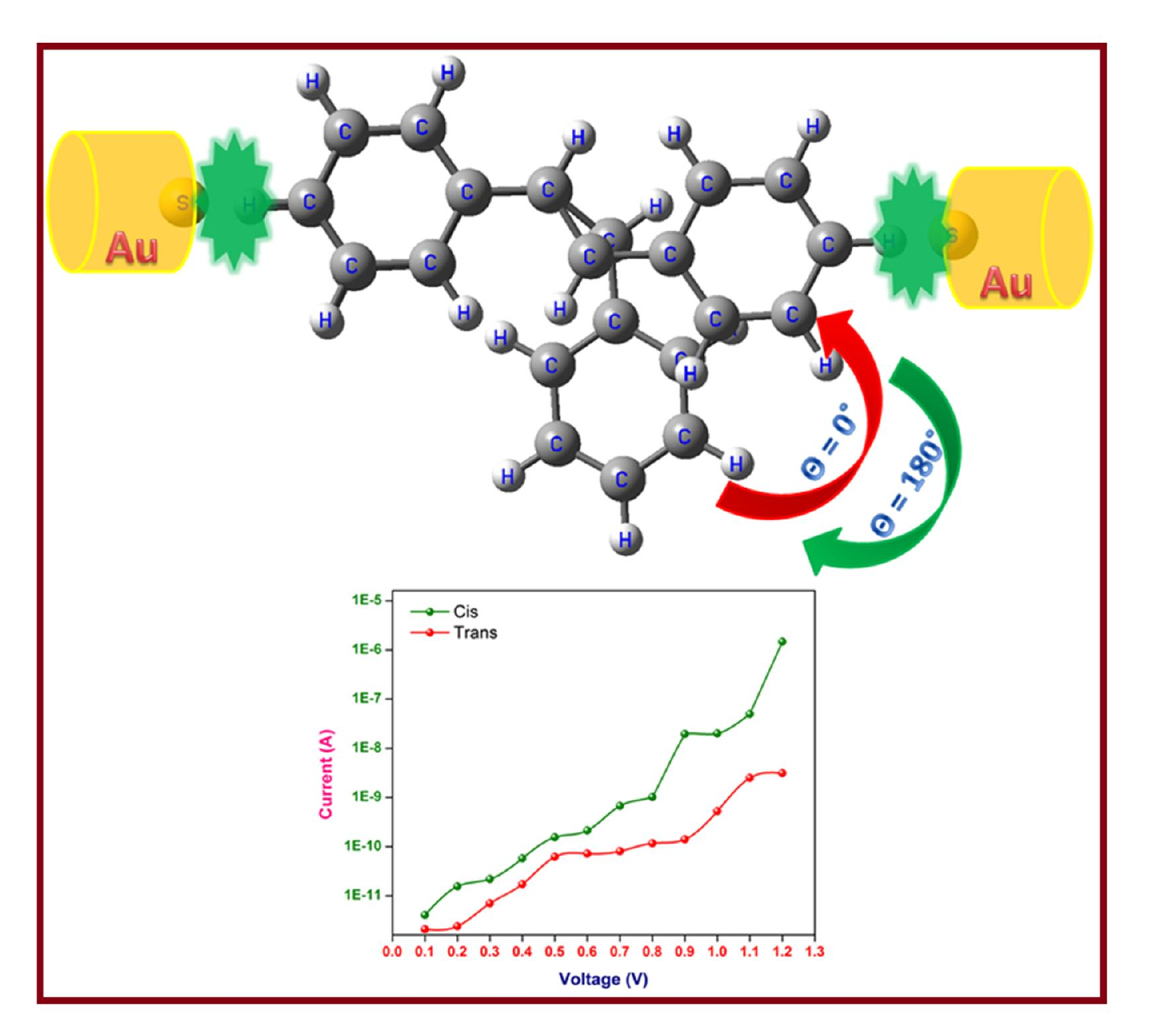}}
\caption{(Colour online) The insights on the switching behaviour of cis and trans stilbene molecular device.}\label{fig-s10}
\end{figure}

As mentioned above, the current for the applied $V_\text{b}$ of stilbene device is given by Landauer-B\"uttiker expression from equation. The current passing over the scattering region in stilbene is noticed to be favorable for cis-SMS device. Besides, the cis$\rightarrow$trans isomerization occurs spontaneously owing to the lack in energy barrier. Nevertheless, substituents in stilbene may convert cis isomer optically more stable than trans isomer upon optical excitation. The current flowing through the cis conformation in stilbene device is comparatively larger than the trans-conformation in stilbene device as noticed in figure~\ref{fig-s9}. The question arises why the trans-isomer has a lower magnitude of current than cis isomer? This is due to the existance of a low energy barrier ($<0.05$~eV) \cite{39} noticed for cis-isomer, whereas the trans-isomer has the barrier of 3~kcal/mol \cite{40}, where the electrons flowing across the scattering region must cross this barrier. Thus, the current flowing through the trans stilbene device is found to be of lower magnitude than cis stilbene device. At low bias voltages (0.1--0.5~V), there is no significant current flow through the device observed both for cis and trans isomer. Beyond 0.5 to 1.2~V, the current transient through the stilbene molecular device is found to be significant. How cis and trans conformers can be used as a switch? From the results of current-voltage characteristics, for cis configuration, the magnitude of current is of the order of microampere for the applied $V_\text{b}$ of 1.2~V. However, the current flowing to trans configuration of stilbene device is only of the order of nanoampere. Thus,  using a proper signal conditioning circuit, the magnitude can be discriminated and can be used as a switch. Moreover, the higher applied bias facilitates the transition across the stilbene scattering region. A drastic variation in the current between the cis and trans configuration can be demonstrated with the help of charge transfer between the HOMO and LUMO levels. Moreover, the variation of current in the stilbene molecular junction is influenced by HOMO-LUMO levels of stilbene conformers and the gold electrodes. Based on the torsion angle $\theta$, the junction barrier differs. As discussed earlier, there is a negligible value of barrier found for the torsion angle $(\theta=180^\circ)$ in cis-isomer, but for trans-isomer, the barrier is observed with a reported value of 3~kcal/mol \cite{40}. Previously we have demonstrated the switching properties of cis-trans isomer of azobenzene molecular device \cite{1}. Leyssner and co-worker \cite{16} demonstrated the molecular switches adsorbed on Au(111) and compared the switching behaviour of azobenzene and stilbene derivatives. This further confirms and validates the switching properties of stilbene with the present research work. Therefore, it is clearly proved that cis and trans-SMS device can be utilized as a switching element. Figure~\ref{fig-s10} gives the insights on the switching behaviour of cis and trans stilbene molecular device.

The stilbene isomerization is mainly utilized for varying supramolecular geometry and domains. Upon visible or ultraviolet irradiation, the cis and trans configuration can be changed in the stilbene device. Thus, we may induce a phase change, a solubility change, surface activity of a surfactant, and the critical micelle concentration can be varied upon shining a particular wavelength of radiation on stilbene molecule. Moreover, it can be recommended that stilbene molecular device should be utilized as a photo-switch as well as may be used for different potential applications including photo-actuation, lithographic patterning and optical switches.

\section {Conclusions}

In conclusion, the switching behaviour of a stilbene molecular system device is investigated using DFT and NEGF technique. The scattering region of stilbene single molecule encompasses cis and trans isomers of stilbene, and its switching properties are investigated. The DDOS spectrum of trans isomer reveals less peak maxima than cis isomer of a stilbene device. By contrast, the Pmax in cis isomer is observed to be high in both the conduction band and in the valence band, which arise owing to the rotation of the torsion angle of C$=$C bond $(\theta=180^\circ)$. The transmission spectrum also proves the presence of a higher peak amplitude in cis-stilbene molecular device (ON state) compared to the trans-stilbene device (OFF state). The current-voltage characteristics reveal a closed configuration of a cis-stilbene molecular device. The magnitude of current flowing though the cis stilbene device is found to be higher than the trans stilbene device. The findings of the proposed research work verify the switching behaviour of SMS device, which is investigated at an atomistic level. Besides, a stilbene based molecular system can have a variety of potential applications including photo-actuation and optical switches.

\section*{Acknowledgements}
The authors wish to express their sincere thanks to Nano Mission Council (No.~SR/NM/NS-1011/2017(G)) Department of Science \& Technology, India for financial support.

\ukrainianpart

\title{Перемикальна поведінка стильбенового молекулярного пристрою: дослідження з перших принципів}%
\author{В. Нагараджан, Р. Чандірамулі}
\address{
Школа електротехніки та електроніки,  університет 
SASTRA, \\ Тірумалайсамудрам, Танджавур --- 613 401, Індія
}

\makeukrtitle

\begin{abstract}
	
Перемикальна поведінка пристроїв, що працюють на основі  стильбенової молекулярної системи (SMS) досліджується з використанням методу нерівноважних функцій Гріна та першопринципних розрахунків. Спектр трансмісії  
 cis-ізомерів підтвердив, що більша кількість електронів  переходить через  SMS-пристрій, використовуючи оптичне збудження спіном зв'язку C$=$C на торсійний кут $(\theta=180^\circ)$. Вольт-амперні характеристики вказують на існування  нижчих  величин струму для trans-стильбенових пристроїв та вищих величин струму для  cis-стильбенових пристроїв при накладанні зовнішньої зміщувальної напруги. В результаті проведеної роботи доведено, що  cis і trans-стильбеновий пристрій можна використовувати в якості перемикача.
\keywords стильбен, молекулярний пристрій, електронна густина, перемикання, трансмісія %
\end{abstract}


\begin{thebibliography}{99}
\bibitem{1}    Dhivya G., Nagarajan V., Chandiramouli R., Chem. Phys. Lett., 2016, \textbf{660}, 27, \doi{10.1016/j.cplett.2016.07.061}.
\bibitem{2}  	 Meier H., Angew. Chem. Int. Ed., 1992, \textbf{31}, 1399, \doi{10.1002/anie.199213993}. 
\bibitem{3}	  Tamai N., Miyasaka H., Chem. Rev., 2000, \textbf{100}, 1875, \doi{10.1021/cr9800816}.
\bibitem{4}    Balzani V., Credi A., Venturi M., ChemPhysChem, 2008, \textbf{9}, 202, \doi{10.1002/cphc.200700528}.
\bibitem{5}	Kaloni T.P., Giesbrecht P.K., Schreckenbach G., Freund M.S., Chem. Mater., 2017, \textbf{29}, 10248,\\ \doi{10.1021/acs.chemmater.7b03035}.
\bibitem{6}	Kaloni T.P., Schreckenbach G., Freund M.S., Sci. Rep., 2016, \textbf{6}, 36554, \doi{10.1038/srep36554}.
\bibitem{7}	Kaloni T.P., Schreckenbach G., Freund M.S., J. Phys. Chem. C, 2015, \textbf{119}, 3979, \doi{10.1021/jp511396n}.
\bibitem{8}	Mehmood U., Al-Ahmed A., Hussein I.A., Renewable Sustainable Energy Rev., 2016, \textbf{57}, 550,\\ \doi{10.1016/j.rser.2015.12.177}. 
\bibitem{9}	Dediu V., Murgia M., Matacotta F.C., Taliani C., Barbanera S., Solid State Commun., 2002, \textbf{122}, 181, \doi{10.1016/S0038-1098(02)00090-X}.
\bibitem{10}	Majumdar S., Laiho R., Laukkanen P., V\"ayrynen I.J., Majumdar H.S., \"Osterbacka R.,  Appl. Phys. Lett., 2006, \textbf{89}, 122114, \doi{10.1063/1.2356463}.
\bibitem{11}	Bouas-Laurent H., D\"urr H., Pure Appl. Chem., 2001, \textbf{73}, 639, \doi{10.1351/pac200173040639}.
\bibitem{12}	Oudar J.L., J. Chem. Phys., 2008, \textbf{67}, 446, \doi{10.1063/1.434888}.
\bibitem{13}	Tian H., Yang S.J., Chem. Soc. Rev., 2004, \textbf{33}, 85, \doi{10.1039/b302356g}.
\bibitem{14}	Lukyanov B.S., Lukyanova M.B., Chem. Heterocycl. Compd., 2005, \textbf{41}, 281, \doi{10.1007/s10593-005-0148-x}.
\bibitem{15}      Klajn R., Chem. Soc. Rev., 2014, \textbf{43}, 148, \doi{10.1039/C3CS60181A}.
\bibitem{16}	Leyssner F., Hagen S., \'Ov\'ari L., Doki\'c J., Saalfrank P., Peters M.V., Hecht S., Klamroth T., Tegeder P., J. Phys. Chem. C, 2010, \textbf{114}, 1231, \doi{10.1021/jp909684x}.
\bibitem{17}   Zhang X., Hou L., Samor\`i P., Nat. Commun., 2016, \textbf{7}, 11118, \doi{10.1038/ncomms11118}.
\bibitem{18}	Guti\'errez R., Grossmann F., Schmidt R., ChemPhysChem, 2003, \textbf{4}, 1252, \doi{10.1002/cphc.200300768}.
\bibitem{19}    Soler J.M., Artacho E., Gale J.D., Garc\'ia A., Junquera J., Ordej\'on P., S\'anchez-Portal D., J. Phys.: Condens. Matter, 2002, \textbf{14}, 2745, \doi{10.1088/0953-8984/14/11/302}.
\bibitem{20}	Taylor J., Guo H., Wang J., Phys. Rev. B, 2001, \textbf{63}, 121104(R), \doi{10.1103/PhysRevB.63.121104}.
\bibitem{21}	Monkhorst H.J.,  Pack J.D., Phys. Rev. B, 1976, \textbf{13}, 5188, \doi{10.1103/PhysRevB.13.5188}.
\bibitem{22}	Perdew J.P., Burke K., Wang Y., Phys. Rev. B, 1996, \textbf{54}, 16533, \doi{10.1103/PhysRevB.54.16533}.
\bibitem{23}	Perdew J.P., Chevary J.A., Vosko S.H., Jackson K.A., Pederson M.R., Singh D.J., Fiolhais C., Phys. Rev. B, 1992, \textbf{46}, 6671, \doi{10.1103/PhysRevB.46.6671}.
\bibitem{24}	Troullier N., Martins J.L., Phys. Rev. B, 1992, \textbf{46}, 1754, \doi{10.1103/PhysRevB.46.1754}.
\bibitem{25}	Bhuvaneswari R., Nagarajan V., Chandiramouli R., Chem. Phys. Lett., 2018, \textbf{691}, 37,\\ \doi{10.1016/j.cplett.2017.10.058}.
\bibitem{26}	Nagarajan V., Dharani S., Chandiramouli R., Comput. Theor. Chem., 2018, \textbf{1125}, 86,\\ \doi{10.1016/j.comptc.2018.01.004}.
\bibitem{27}	 Nagarajan V., Bhattacharyya A., Chandiramouli R., J. Mol. Graphics Modell., 2018, \textbf{79}, 149,\\ \doi{10.1016/j.jmgm.2017.11.009}.
\bibitem{28}	Pomorski P., Roland C., Guo H., Phys. Rev. B, 2004, \textbf{70}, 115408, \doi{10.1103/PhysRevB.70.115408}.
\bibitem{29}	Brandbyge M., Mozos J.-L., Ordej\'on P., Taylor J., Stokbro K., Phys. Rev. B, 2002, \textbf{65}, 165401, \doi{10.1103/PhysRevB.65.165401}.
\bibitem{30}	Bhuvaneswari R., Nagarajan V., Chandiramouli R., Chem. Phys., 2018, \textbf{501}, 78,\\ \doi{10.1016/j.chemphys.2017.12.003}.
\bibitem{31}	Nagarajan V., Chandiramouli R., IEEE Sens. J., 2018, \textbf{18}, No.~3, 948, \doi{10.1109/JSEN.2017.2781728}.
\bibitem{32}	Evans C.H., Reynisson J., Geirsson J.K.F., Kvaran \'A., McGimpsey W.G., J. Photochem. Photobiol., A, 1998, \textbf{115}, 57, \doi{10.1016/S1010-6030(98)00243-3}.
\bibitem{33}	Hub W., Schneider S., Doerr F., Oxman J.D., Lewis F.D., J. Am. Chem. Soc., 1984, \textbf{106}, 701,\\ \doi{10.1021/ja00315a040}.
\bibitem{34}	Fu{\ss} W., Kosmidis C., Schmid W.E., Trushin S.A., Angew. Chem. Int. Ed., 2004, \textbf{43}, 4178,\\ \doi{10.1002/anie.200454221}. 
\bibitem{35} 	Waldeck D.H., Chem. Rev., 1991, \textbf{91}, 415, \doi{10.1021/cr00003a007}.
\bibitem{36}	Baskin J.S., Ba\~nares L., Pedersen S., Zewail A.H., J. Phys. Chem., 1996, \textbf{100}, 11920, \doi{10.1021/jp960909x}.
\bibitem{37}	Fu{\ss} W., Kosmidis C., Schmid W.E., Trushin S.A., Chem. Phys. Lett., 2004, \textbf{385}, 423,\\ \doi{10.1016/j.cplett.2003.12.114}.
\bibitem{38}	Syage J.A., Felker P.M., Zewail A.H., J. Chem. Phys., 1998, \textbf{81}, 4706, \doi{10.1063/1.447520}.
\bibitem{39}	Abrash S., Repinec S., Hochstrasser R.M., J. Chem. Phys., 1998, \textbf{93}, 1041, \doi{10.1063/1.459168}.
\bibitem{40}	Quenneville J., Mart\'inez T.J., J. Phys. Chem. A, 2003, \textbf{107}, 829, \doi{10.1021/jp021210w}.  
\bibitem{41}	Deekshitha M., Nagarajan V., Chandiramouli R., Chem. Phys. Lett., 2015, \textbf{641}, 129,\\ \doi{10.1016/j.cplett.2015.10.070}.
\bibitem{42}	Nagarajan V., Chandiramouli R., Solid State Commun., 2018, \textbf{269}, 50, \doi{10.1016/j.ssc.2017.09.023}.
\bibitem{43}	Nagarajan V., Chandiramouli R., Comput. Theor. Chem., 2017, \textbf{1105}, 52, \doi{10.1016/j.comptc.2017.02.023}.
\bibitem{44}	B\"uttiker M., Imry Y., Landauer R., Pinhas S., Phys. Rev. B, 1985, \textbf{31}, 6207, \doi{10.1103/PhysRevB.31.6207}.
\bibitem{45}	Stokbro K., J. Phys.: Condens. Matter, 2008, \textbf{20}, 064216, \doi{10.1088/0953-8984/20/6/064216}.
\bibitem{46}	Brandbyge M., S\o{}rensen M.R., Jacobsen K.W., Phys. Rev. B, 1997, \textbf{56}, 14956, \doi{10.1103/PhysRevB.56.14956}.
\bibitem{47}	Taylor J., Guo H., Wang J., Phys. Rev. B, 2001, \textbf{63}, 245407, \doi{10.1103/PhysRevB.63.245407}. 
\bibitem{48}	Chandiramouli R., Nagarajan V., J. Comput. Electron., 2017, \textbf{16}, 316, \doi{10.1007/s10825-017-0956-0}.  
\bibitem{49}	Nagarajan V., Chandiramouli R., IEEE Trans. Nanotechnol., 2017, \textbf{16}, 445, \doi{10.1109/TNANO.2017.2682125}.  
\bibitem{50}	Nagarajan V., Chandiramouli R., Chem. Phys. Lett., 2017, \textbf{675}, 131, \doi{10.1016/j.cplett.2017.03.031}.
\bibitem{51}	Nagarajan V., Dhivya G., Chandiramouli R., J. Comput. Electron., 2018, \textbf{17}, 1, \doi{10.1007/s10825-017-1047-y}.  
\bibitem{52}	Zhou Y.H., Zheng X.H., Xu Y., Zeng Z.Y., J. Chem. Phys., 2006, \textbf{125}, 244701, \doi{10.1063/1.2409689}.
\end{thebibliography}
\end{document}